# Detection and Prevention Against RTS Attacks in Wireless LANs

Tauseef Jamal, Muhammad Mussadiq Umair, M. Alam and Zeeshan Haider
CoCoLab, PIEAS Islamabad.
saddi1991@live.com, luckier19@gmail.com

*This is the author preprint version. This work is supported by CoCo Lab under project, "Smart Security Framework for MANets". Original work is presented by Mr. Musadiq Umair as MS thesis in Computer Science from PIEAS, in 2017 an extract from his work has been published in IEEE CCode, while extension of work was carried by other student Kirmat Ullah in his MS thesis in 2017. His work was published in IARIA ICDS 2018. This paper mainly explains the novel MAC layer of IEEE 802.11, which is the real brain of whole project. For full paper refer to https://ieeexplore.ieee.org/document/7918920/*

*Abstract*— Widely deployed wireless network devices use a shared medium to communicate among mobile nodes. WLAN uses virtual carrier sensing mechanism to solve issues like hidden node problem, while this mechanism is vulnerable to DOS attacks e.g., RTS attack. This paper discusses RTS (Request to Send) attack where malicious nodes reserve the medium unnecessarily for overdue period of time. The effects of such attacks on performance of WLAN are observed, proposed a mitigation technique to restore network performance and analyzed the improvement in performance.

## I. INTRODUCTION

Wireless communication is the fast growing industry and will continue to evolve. Wireless technology provides a flexible data communications system that uses radio frequency to transmit and receive information over the air instead of physical wires. Wireless connectivity allows free movement as users are no longer bound by wires and can access network resources anywhere within coverage range of network. Such network provides reliability, scalability, flexibility and rapid deployment. Wireless networks are been widely used in Cellular networks, WLANs, WSN, Satellite communication etc.

Security is highly considered factor of networks today, it is the protection of network resources and data transferred over the network. There are numerous factors that can affect Confidentiality, Integrity and Availability of a system. In order to protect the network from data loss, disruption or exposure certain procedures like access control, encryption, authentication, redundancy etc. are applied to improve the performance of network. Analyzing the performance metrics, availability is particularly important to keep a network's performance. There are several possible Denial of Service attacks that can cause the disruption and decrease the network bandwidth & throughput by increases the delays & data errors. In such attacks it becomes challenging to track and refuse attacker's request without refusing legitimate requests.

A WLAN is a network that links two or more devices using a wireless medium, so that devices can communicate to each other within a limited range, depending on architecture of network. WLANs use either Infrastructure (centrally controlled) or Ad-hoc (independent) to communicate among wireless devices. The shared medium in wireless network raised issues like collision and interference which leads to higher data losses, low throughput, high latency and congestion that can be lead to failure.

WLAN uses virtual carrier sensing mechanism to solve issues like hidden node problem, while this mechanism is vulnerable to DOS attacks e.g., RTS attack. In this mechanism, the node that want to communicate send an RTS frame holding reservation duration field (time for which the nodes are expecting to complete their data transfer). As soon as destined node receives the RTS, it responds with a CTS frame containing the remaining duration. Every other node that receives either RTS or CTS updates their Network Allocation Table (NAV), where NAV is a timer that can uniformly reduce to zero. The nodes other than sender/receiver are not allowed to transmit unless there NAV is zero, hence stay quiet for duration of the other conversation. After all set up, the node sends DATA and waits for ACK, and completes the process. As shown in Figure 1.

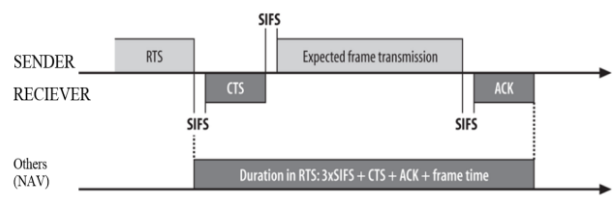

**Figure 1: Virtual Carrier Sensing Mechanism [1]**

This paper presented a lightweight solution for RTS attack's identification and mitigation. This solution is based on re-evaluation and implicit notifications in order to isolate malicious node.

## II. RTS ATTACK

WLAN use control frames with 2 bytes for duration field which can be set as required to communicate and represents duration in microseconds, the standard allows maximum duration of 32,767us. A node with malicious intent can modify the RTS control frame's set duration operation and replace it with a larger value to pretend that it needs large duration to communicate, to held the channel unnecessarily for overdue time and reduce the bandwidth utilization with increase in latency for other communications (c.f. Figure 2)

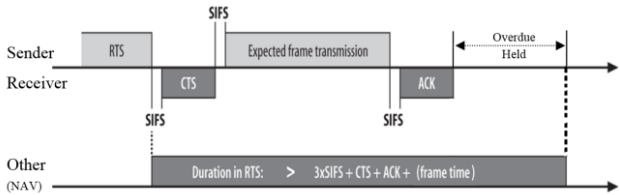

**Figure 2: RTS Attack behavior**

In case of RTS attack the devices in communication remain in hold of channel event after the communication is over, they can initiate another communication and enforce new large NAV value and keep the hold forever.

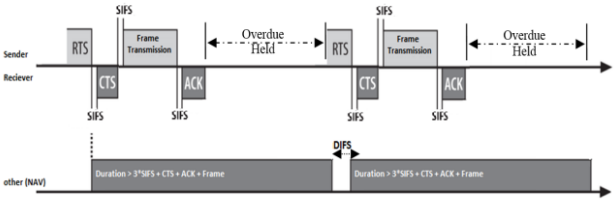

**Figure 3: RTS Attack with RTS Flood**

I what concerns the prior work, most of the work is relying over periodic broadcast and/or explicit messages. For effective mitigation such broadcast must be very frequent, which increases the overhead. Following describe some of the related work.

### A. RELATED WORK

Authors in [2] suggested and examine a statistical approach to detect the NAV attack in WLAN on MAC layer. Since all node return CTS in response to RTS, the attacker can assert large duration field as NAV to not let other nodes to access the medium, named the problem as NAV attack. The authors used statistical information (mean and standard deviation) of network to detect the attack. However, the malicious user can exploit it, by maintaining statistics and trying to keep its states like normal nodes.

In [3], authors discussed congestion in RTS/CTS where nodes hold channel with RTS, and raise false blocking issue, therefore, no node will be able to use medium as they think channel is busy.

Similarly, work in [4] analyzed Ad-Hoc network, where DCF is vulnerable to Denial of Service attacks. The attacker sends a control frame (RTS, CTS or DATA) to a non-existing node, while all overhearing nodes defer their access. They discussed the effects of these attacks and determine that performance is highly disturbed by launching RTS attack. Especially, in case when RTS is sent to Access Point (AP), the range of attack increases a lot.

Author in [5] explored flaws of WLAN, where the disobedient device can stall adjacent devices from gaining access to medium for long duration. They discussed potential virtual jamming attacks like NAV attacks and spurious CTS, a backward compatible solution to avoid NAV attack was suggested to amend IEEE 802.11, to have ability to reset the NAV if the channel is found idle for long. The solution was NAV validation to overcome these weaknesses, the idea is to set extra MAC layer timers: RTS_DATAHEAD_Time and CTS_ACK_Time. RTS_DATAHEAD_Time keeps track of time between RTS and DATA packet, while CTS_ACK_Time keeps track of the period between CTS and ACK packet. The timer checks if the DATA and ACK packets are received as expected otherwise reset NAV to reset the medium. Since all devices will be maintaining the timers and continually sensing the channel they will consume a lot of energy and increase in overhead.

In [6] authors studied the virtual carrier sense attacks and their practical effectiveness, where attacker transmits packets with diverse large duration values. The paper focused the research to study WLAN performance under RTS flooding Denial of Service attack, under several conditions such as no. of attackers, retransmissions count and control frames. However, the proposed solution only works in non-hidden scenarios where node can determine if CTS is for a valid RTS or not.

While [7] discussed and exploits various attacks caused by control frames RTS, CTS and ACK frames. The attacks (Virtual Carrier Sense Flooding) are being established by manipulating there duration field. They carried numerous experiments to observe the effect of such attacks on wireless network. They established the fact that VCSF attacks can affect network more than those on the APs, also that the attacks using RTS are more disastrous than any other control frame.

The author of [8] discussed spurious CTS attack (SCTS) and presented Channel Sensing based Discarding (CSD) mechanism where nodes consume all energy by keep sensing the medium after regular period.

Hence, we proposed a solution to help to determine malicious intent and prevent the attack without use of any additional control messages, periodic polling and periodic broadcasts as well as maintaining table. This way we can avoid unnecessary overhead of additional messages and broad cast, as well as save energy due to avoid unnecessary sensing.

III. PROPOSED SOLUTION

When a node receives an RTS frame from other node, it can store the duration field that has come along with it. Send CTS as expected, with remaining duration mentioned in its duration field. As soon as DATA packet arrives at receiver node it can be analyzed to detect the malicious intent of sender if any. The receiver than evaluates the duration required for the length of this Data frame. Comparing the previously stored RTS duration and the re-evaluated duration of Data the receiver knows if the frame received is legitimate or malicious. The scheme is named RRD (Re-Evaluate RTS Duration). RRD Algorithm for RTS attack's detection is described below:

If Received Frame Type = RTS
       LastRTSDuration ← Frame Duration
Send CTS Frame
If Received Frame Type = DATA
If (LastRTSDuration – 3*SIFS – Duration CTS – Duration ACK) >
       (Duration require of Frame Length
             or Duration field of Data Frame)
       Then Frame is Malicious

Figure 4, shows RRD flowchart for attack detection.

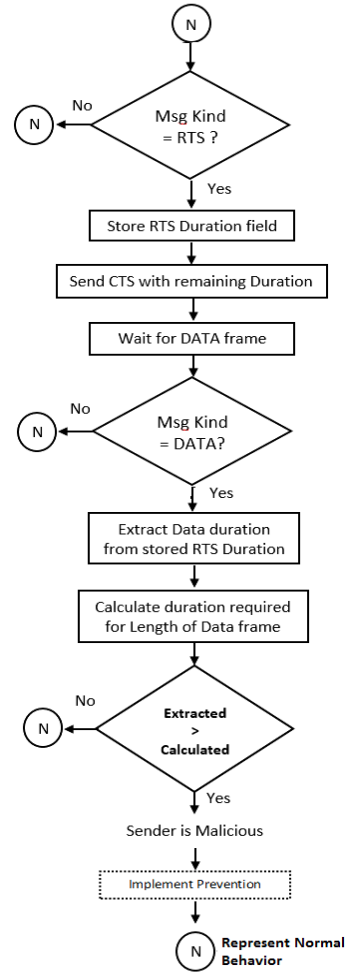

**Figure 4: Flow chart for Detection of RTS Attack**

*A. IMPROVEMENT IN DETECTION SCHEME*

As CTS sent in response to RTS, it calculates remaining duration. So we can store expected data duration by subtracting duration require for ACK. As soon as Data frame arrives its duration field can be compared to the stored expected duration, and determine if frame is malicious. This scheme seems to have the lower overhead in comparison to previously discussed steps. Algorithm of improved detection is described below:

If Frame Received kind = RTS
Send CTS Frame
       Calculates the Remaining Duration
             Expected_Data_Duration ← Remaining_duration
                       - ACK_duration
If Frame Received kind = DATA
If (Expected_Data_Duration > Duration field of Data Frame)
       Then Frame is Malicious

Figure 5, shows RRD flowchart for improved attack detection.

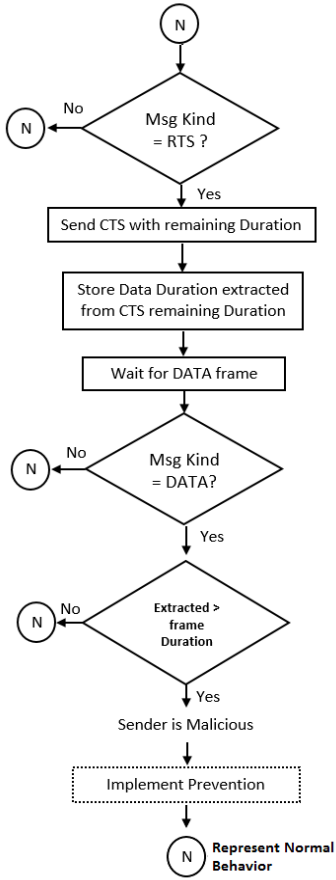

**Figure 5: Flow chart for improved detection**

*B. PREVENT AGAINST ATTACK*

The next important step is to protect network and stop such attacks further. We can implement prevention against RTS attack in two phases:

Phase 1: If an RTS frame received from a node is detected malicious, we add the malicious user information (i.e., MAC address) to a list. To do so nodes should have ability to maintain Black-List, then don't respond any RTS if received from a node in Black-List, instead the nodes will begin new cycle according to legacy protocol (c.f. Figure 6).

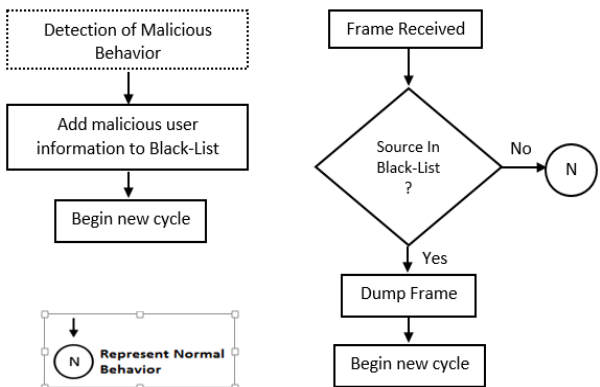

**Figure 6: Prevention against RTS Attack – Phase 1**

So, this node may no longer involve itself in communication with malicious node. It will increase the chances of other nodes to utilize the medium, increasing the throughput of network and decrease in latency.

Phase 2: As soon as we detect a malicious intent of a node, node at receiving end can transmit information to other nodes to update their black-list so that they don't update there NAV in case of frames received from a black-list node. This is done via sending a broadcast ACK frame, including malicious node MAC address in address 3 field. This ACK message plays a role of implicit notification for other nodes to releases their NAV and updates local black list.

Since the malicious node has dedicated medium for conversation with another node for a long duration, it can send another RTS with long duration and since it may be greater than the current NAVs of other nodes so they will have to update there NAVs causing them not to access medium for communication. Therefore by overhearing ACK frame immediately after an RTS, the nodes will reset their NAV counters.

All nodes maintain a black list locally and updated via implicit ACK control message, this way the network can be protected by disallowing attacker to use other node in network, eliminating all packet from those who have been detected for malicious behavior. Unlike prior work, the local black list is never broadcasted, thus, reducing the overhead. Figure 7 illustrates the idea of phase 2 presentation.

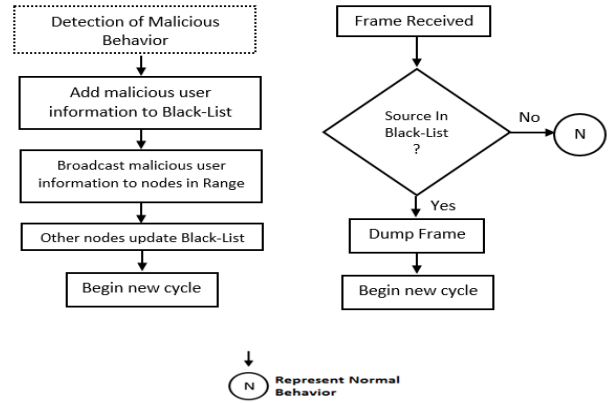

**Figure 7: Prevention against RTS Attack – Phase 2**

## IV. EVALUATION:

Evaluation is based on simulations run on the MiXiM Framework, based OMNet++ discrete event simulator V5.0 using windown plateform [9]. Since, MiXiM has implemneted MAC and Physical layers in details [10]. Therefore, we have exploited the Physical layer parameters using MAC frames and channel access. Table I lists the simulation parameters. Each simulation has a duration of 500 seconds and consider a scenario where all nodes are static.

We investigated a typical WLAN senario composed of one AP (destination) and 25 nodes distributed randomly, where each node is a source. Each node is allocate a unique MAC address, and one wireless channel is shared among all nodes.

We have used thourghput and latencty as performance parameters. All simulation were run with varying number of nodes (2-25).

Table I
SIMULATION PARAMETERS

| Parameter | Values |
|---|---|
| Playground Size | $200 \times 200 m^2$ |
| Path Loss Coefficient | 4 |
| Carrier Frequency | 2.412e9 Hz |
| Max Transmission Power | 100 mW |
| Signal Attenuation Threshold | -120 dBm |
| MAC Header Length | 272 bits |
| MAC Queue Length | 14 frames |
| Basic Bitrate | 1 Mbps |
| Rts-Cts Threshold | 400 bytes |
| Thermal Noise | -110 dBm |
| MAC Neighborhood Max Age | 100 s |
| Payload Size | 1 K bytes |

Figure 8 shows the latency analysis for network with one malicious node. Due to malicious node the latency increases, however, our proposal achieves significant improvement. The gain is higher when density is high, due to more chances of communication.

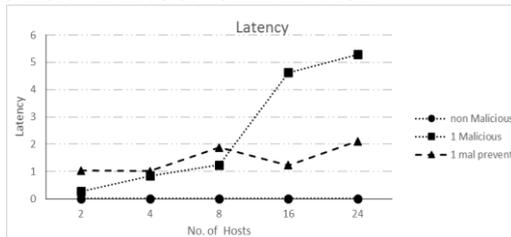

**Figure 8: Latency Vs Density**

As a result latency is slightly higher for less number of nodes, because the overhead get dominating.

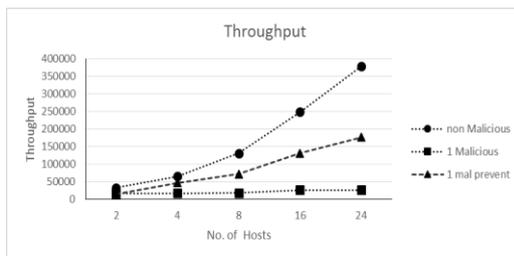

**Figure 9: Throughput Vs Density**

The throughput has significantly increased, due to ACK broadcast mechanism to allow other overhearing nodes to reject NAV updating for black-listed users. Therefore, only allowed nodes are communicating and malicious nodes are automatically isolated.

## V. CONCLUSION AND FUTURE WORK:

Increasing usage of wireless has raised its importance and hence the network performance is a very important factor that needs to be maintained and improved. Wireless networks are vulnerable to Denial of Service attacks. This paper discussed RTS attack that affects the Ad-hoc wireless networks exploiting virtual carrier sensing mechanism that controls medium access for wireless station.

We presented a solution based on redetermination of required duration mechanism to detect the attack, and prevent overhearing node to avoid attacker. This helps restore network performance.

In future, looking forward to improve the prevention scheme by maintain attackers history and make decisions based on previous behavior so that no legitimate user get illuminated for false detection. This way we can extend the work to be incorporated in cooperative network using Relay Spot [11-12]. Hence, we will check how cooperation and network coding can affect our proposed solution. Then the solution will be implemented to real testbed and results can be analyzed to improve the performance.

REFERENCES:

[1] T. Jamal, "Design and Performance of Wireless Cooperative Relaying," PhD. thesis, Univ. of Aveiro, Aveiro, Portugal, 2013.

[2] T. Jamal, and P. Mendes, "COOPERATIVE RELAYING FOR DYNAMIC NETWORKS", EU PATENT, (EP13182366.8), Aug. 2013.

[3] B. Chen, V. Muthukkumarasamy, N. Guimaraes, P. Isaias, and A. Goikoetxea, "Denial of service attacks against 802.11 DCF," in Proceedings of the IADIS International Conference: Applied Computing, 2006.

[4] M. Malekzadeh, A. A. A. Ghani, S. Subramaniam, and J. Desa, "Validating Reliability of OMNeT in Wireless Networks DoS Attacks: Simulation vs. Testbed," International Journal of Network Security, vol. 3, pp. 13-21, 2011.

[5] T. Jamal, P. Mendes and A. Zuquete. "Analysis of hybrid relaying in cooperative WLAN", In proc. IFIP/IEEE Wireless Days, 2013.

[6] D. Chen, J. Deng, and P. K. Varshney, "Protecting wireless networks against a denial of service attack based on virtual jamming," in ACM MobiCom, 2003, pp. 14-19

[7] J. Bellardo and S. Savage, "802.11 Denial-of-Service Attacks: Real Vulnerabilities and Practical Solutions," in USENIX security, 2003, pp. 15-28.

[8] T. Jamal, P. Mendes, and A. Zúquete, "Interference-Aware Opportunistic Relay Selection", In Proc. of ACM CoNEXT student workshop, Tokyo, Japan, Dec. 2011.

[9] T. Jamal, and P. Mendes, "RelaySpot, OMNET++ Module", Software Simulator Extension In Proc. of COPE-SW-13-05, 2013.

[10] T. Jamal and P. Mendes, "802.11 Medium Access Control In MiXiM," Tech Rep. SITILabs-TR-13-02, University Lusófona, Mar. 2013

[11] T. Jamal and P. Mendes, Cooperative Relaying in Dynamic Wireless Networks under Interference Conditions (2014), in: IEEE Communication Magazine, Special issue on User-centric Networking and Services.

[12] T. Jamal, P. Mendes, and A. Zúquete, "Wireless Cooperative Relaying Based on Opportunistic Relay Selection," International Journal on Advances in Networks and Services, vol. 05, no. 2, pp. 116–127, Jun. 2012.